\documentstyle[prl,aps,epsf]{revtex}
\draft
\begin{document}
\twocolumn[\hsize\textwidth\columnwidth\hsize\csname @twocolumnfalse\endcsname

\title{Ordering and Melting in Colloidal Molecular Crystal Mixtures 
} 
\author{C. Reichhardt and C.J. Olson Reichhardt} 
\address{ 
Center for Nonlinear Studies and Theoretical Division,
Los Alamos National Laboratory, Los Alamos, NM 87545}

\date{\today}
\maketitle
\begin{abstract}
We show in simulations that 
a rich variety of novel orderings such as pinwheel and star states can be 
realized for colloidal molecular crystal mixtures at
rational ratios of
the number of colloids to the number of minima from an
underlying periodic substrate. 
These states can have multi-step melting transitions and also 
show coexistence 
in which one species disorders while 
the other species remains orientationally ordered. 
For other mixtures, only partially ordered or frustrated states form.    
\end{abstract}
\vspace{-0.1in}
\vspace{-0.3in}

\vskip2pc]
\narrowtext
Recently a 
new type of colloidal state termed colloidal molecular crystals (CMCs)
for colloids interacting with two-dimensional periodic substrates 
has been proposed in simulations \cite{Reichhardt,Olson}, 
observed experimentally \cite{Bechinger}, 
and studied theoretically \cite{Agra,Sarlah}.  
CMCs occur
when the  number of colloids is an integer multiple $n$ of the number of
substrate minima, and the 
$n$ colloids
residing in each minimum form a composite $n$-mer object  
such as a dimer or trimer.
Neighboring $n$-mers can have an orientational
ordering relative to one another in addition to the spatially periodic ordering
imposed by the underlying substrate.
The CMCs were named in analogy 
with molecular crystals, in which the molecules 
have orientational as well as crystalline ordering.

CMCs exhibit interesting two-step melting transitions.
The first stage of melting occurs when the $n$-mers 
begin to rotate and lose orientational ordering, while in 
the second stage, the
$n$-mers dissolve and individual colloids
begin to diffuse throughout the sample \cite{Reichhardt,Bechinger}. 
The orientational melting was shown to fall into 
an Ising-like transition class \cite{Agra,Sarlah}.  
A reentrant disordering can also occur 
in which an orientationally ordered $n$-mer state 
disorders when the substrate strength is increased \cite{Olson,Bechinger}.
Recent theoretical work 
has shown that
the orientational ordering arises due to the 
anisotropic multipole interactions between the $n$-mers \cite{Agra}.
The reentrant disordering occurs 
when the $n$-mer is compressed by the increasing substrate force,
decreasing both the distance between the colloids and 
the multipole moment, until
the multipole interaction energy drops below the thermal energy.

When the substrate is varied, new states emerge.  Studies of colloids
interacting with arrays of small traps embedded in a smooth background have
shown coexistence between an
interstitial colloidal liquid and a pinned ordered solid \cite{Leiderer}. 
In the case of colloids driven over 
two-dimensional (2D) substrates, interesting
locking effects occur when the colloids move preferentially along certain
symmetry directions of the underlying 
lattice \cite{Grier,Spalding,Gopinathan,EPL}.     
Colloidal crystallization on periodic substrates may prove to
be a useful method to create novel colloidal structures for
photonic band gap device or filter applications \cite{Review}.  
Additionally, colloids
are an ideal model system for
studying collective particle states as well as topological defects generated
on periodic substrates, which is relevant  
to a variety of other systems 
including atoms and molecules adsorbed on surfaces 
\cite{Berlinsky,Goege}, 
superconducting vortices interacting with 
periodic pinning sites \cite{Baret}, or vortices in
Bose-Einstein condensates interacting with periodic optical
traps \cite{Duine}.  
 
An open question is what kind of ordering and melting transitions
occur for CMCs that are not pure, but are composed of
a mixture of $n$-mers and $m$-mers, where $m=n+1$.
In this work we show that when the number of colloids $N_c$ is
a rational, noninteger multiple $p$ of the number of substrate minima $N_s$, 
a variety of novel crystalline and 
partially crystalline colloidal states form which cannot
occur at integer fillings. These states include superlattices of
$n$- and $m$-mer mixtures for certain values of $p$.
As one example, we demonstrate pinwheel 
ordering similar to that observed for molecular dimers adsorbed on 
2D atomic substrates. 
Other orderings, such as
star shapes in which dimers point to a monomer, are also possible. 
These mixed states show a rich multi-step melting behavior
in which the $n$-mers disorder one species at a time, 
followed by the dissolution of all of the $n$-mer states.
The partially crystalline fillings may offer a convenient realization of
an orientational glass \cite{Binder}.

We consider a  2D system of $N_c$ colloids with 
periodic boundary conditions in the $x$ and
$y$ directions, and employ 
Langevin dynamics as used in previous 
CMC simulations \cite{Reichhardt}.
The equation of motion for 
a colloid $i$ is
\begin{equation}
\frac{d {\bf r}_{i}}{dt} = {\bf f}_{i}^{cc} + {\bf f}_{s} + {\bf f}_{T} \ .
\end{equation}
Here ${\bf f}_{i}^{cc} = -\sum_{j \neq i}^{N_{c}}\nabla_i V(r_{ij})$ 
is the interaction force from the other colloids, where
we assume
a screened Coulomb interaction,
$V(r_{ij}) = (Q/|{\bf r}_{i} - {\bf r}_{j}|)
\exp(-\kappa|{\bf r}_{i} - {\bf r}_{j}|)$. $Q=1$ is the charge of the 
particles, $1/\kappa$ is the screening length, 
and ${\bf r}_{i(j)}$ is the position of
particle $i(j)$.  
The substrate force 
comes from a triangular substrate such as that used 
in recent experiments \cite{Bechinger}, and is given by 
${\bf f}_{s} = \sum_{i=1}^{3}A\sin(2\pi b_i /a_{0})
[\cos(\theta_{i}){\hat {\bf x}} - \sin(\theta_{i}){\hat {\bf y}}]$,
where $b_i=x \cos(\theta_{i})-y\sin(\theta_{i}) + a_0/2$, 
$\theta_{1} = \pi/6$, $\theta_{2} = \pi/2$ and $\theta_{3} = 5\pi/5$.   
Here $A$ is the strength of the substrate. 
We measure length in units of the
fixed substrate lattice constant $a_{0}$, and
take $\kappa = 2/a_{0}$.
We have also considered the case of square substrates.
The thermal force ${\bf f}_{T}$ is a 

\begin{figure}
\center{
\epsfxsize=3.5in
\epsfbox{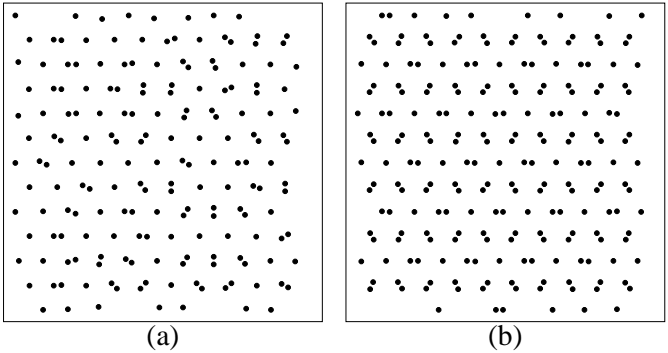}}
\caption{
The colloid configurations (black dots) at $T = 0$ for
a triangular 2D periodic substrate with $A = 2.5$. 
(a) A filling of $N_c/N_s=3/2$, where a disordered
or frustrated configuration occurs.    
(b) A filling of $N_c/N_s=7/4$, where a star ordering occurs. 
}
\end{figure}

\noindent
randomly fluctuating force 
from random kicks.  
To obtain initial configurations, we start the system at a temperature 
where all the colloids are diffusing rapidly, and
gradually cool to $T = 0$. 
In this model, we do not take
into account hydrodynamic effects or possible long-range attractions between
colloids.
We focus on the regime where $A$ is sufficiently large that 
the colloids are easily trapped. In the optical trap experiments, $A$ can be 
readily tuned by varying the laser strength.
We have checked our results for different system sizes
and screening lengths and find no qualitative changes.
To characterize the phases, we measure the finite time particle 
displacements 
$\delta r = (1/N_c)\sum_{i=1}^{N_c}|{\bf r}_i(t) - {\bf r}_i(t^\prime)|$, 
where $t^{\prime} < t$,
as well as the fraction of six-fold coordinated colloids $P_6$.  

We first consider mixtures of monomer and dimer colloid
states, $1 < N_{c}/N_{s} <  2$. 
Previous work has shown that a herringbone structure is
stabilized at $N_{c}/N_{s} = 2$ (see Fig. 2(b) of Ref.~\cite{Reichhardt}).  
At $N_{c}/N_{s} = 3/2$, illustrated in Fig.~1(a), 
we find a disordered arrangement of dimers and monomers. 
The system is frustrated since it is not possible
to tile the triangular lattice with an alternating arrangement of
monomers and dimers.  Thus the monomer and dimer locations
are disordered, and in addition, the dimers have only
local orientational ordering.
For $N_c/N_s$ close to and including 3/2, 
the freezing temperature for dimer rotations
is very low due to the lack of orientational order.
In contrast, at $N_c/N_s=7/4$ we find an ordered 
state which we refer to as
{\it star} ordering, shown in Fig.~1(b).
Every other 
minimum in every other row captures a monomer while the remaining
sites are occupied by dimers. All the  dimers align such that
they point toward the monomers, producing starlike structures.
For $N_c/N_s=7/4$ 
the orientational disordering temperature, where
the dimer orientational ordering is lost but the colloids remain localized, 
closely coincides with the 
orientational disordering temperature $T_{c}$ of the herringbone state 
for $N_{c}/N_{s} = 2$.    

\begin{figure}
\center{
\epsfxsize=3.5in
\epsfbox{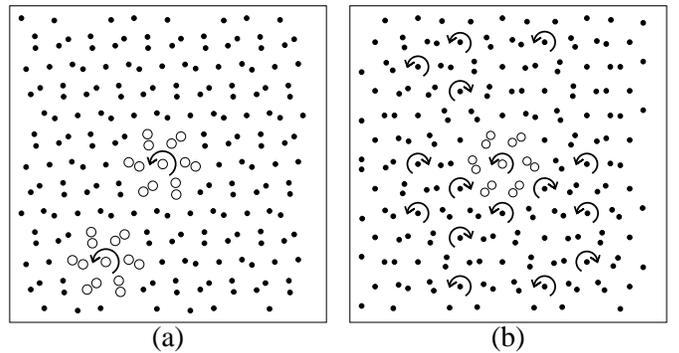}}
\caption{ 
(a) Colloid configuration
(black dots) for filling of $N_c/N_s=7/4$ at $A = 1.5$, where
a pinwheel state occurs. Two example pinwheels are highlighted with white
circles and the rotational direction is marked with an arrow.
(b) Colloid configuration for an orientationally 
frustrated filling of $N_c/N_s=5/3$ at $A = 2.5$.
Local pinwheel ordering is indicated by the arrows.
An example distorted pinwheel is highlighted with white circles.
}
\end{figure}

The star state shown in Fig.~1(b) only occurs for sufficiently large
$A > 2.0$.  When the substrate is weaker, for $0.5 < A < 2.0$, 
a pinwheel ordering occurs
instead, in which the dimers surrounding the monomer have 
an additional tilt as illustrated in
Fig.~2(a).  There are two possible directions for this tilt and Fig. 2(a)
shows the counterclockwise case; for different initial random seeds there
is a 50\% probability of producing a clockwise arrangement instead.
Pinwheel states
are predicted for the adsorption of dimer or linear molecules
on triangular substrates at densities such that a portion of the dimers 
stand perpendicular to the surface, surrounded 
by the remaining dimers which lie parallel to the surface 
\cite{Berlinsky,Goege}.
They can also occur for adsorbed atom-molecule mixtures such as
CO$_{(1-x)}$Ar$_x$ \cite{mixtures}.
Pinwheel states with a four-molecule basis are expected for a filling of $7/4$ 
\cite{Berlinsky}, and 
pinwheel states with a larger basis are also possible for 
higher fillings
such as 
$13/7$ \cite{Goege}. 
For $ A < 0.5$ at $N_c/N_s=7/4$,
the colloid-colloid interactions begin to dominate
over the substrate interaction and 
the colloids form an incommensurate modulated triangular
lattice.  

By frustrating the system, we can create an orientationally disordered
state.  For example, 
in Fig.~2(b) we show a partially ordered state that
occurs at a filling of $N_c/N_s=5/3$. 
Here, unlike the case in Fig.~1(a), there is positional ordering, 
since every third site in every
row contains a monomer.
The dimers, however, 
do {\it not} form an orientationally ordered arrangement, although
locally the dimers have pinwheel or distorted pinwheel structures.
Both chiralities of the pinwheels appear, as indicated
in Fig.~2(b), but
there is no long range dimer orientational  ordering. 
At fillings $N_c/N_s=5/4$  and $N_c/N_s=4/3$, the 
dimers form a triangular lattice;
however, since the spacing between neighboring dimers is larger 
than $a_0$ in these configurations, 
the 

\begin{figure}
\center{
\epsfxsize=3.5in
\epsfbox{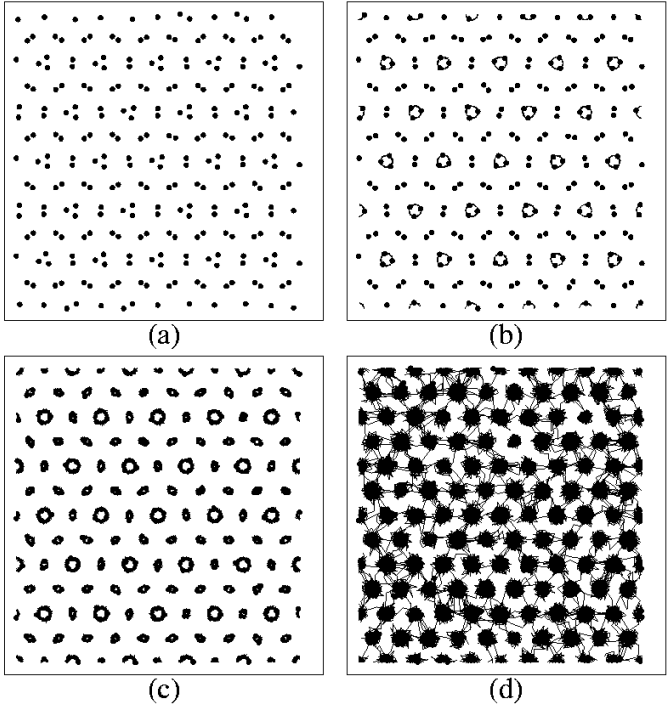}}
\caption{
The colloidal positions (black dots) and trajectories (lines) over a
fixed period of time
for $A = 3.0$ and a filling of $N_c/N_s=9/4$ at different
temperatures. 
(a) Phase I, trimer-dimer solid, at $T/T_{m} = 0$.  
(b) Phase II, trimer liquid, dimer solid phase, at $T/T_{m} = 0.56$. 
(c) Phase III, trimer-dimer liquid, at $T/T_{m} =2.4$. (d) 
Modulated liquid phase, at $T/T_{m} = 5.0$.      
}
\end{figure}

\noindent
quadrupole-quadrupole
interaction strength is too small to impose orientational order
and the dimers are rotationally disordered even at very low temperatures.     

For fillings $2 < N_c/N_s < 3$, we obtain mixtures of dimers and
trimers.
At a filling of $N_c/N_s=5/2$ the dimer-trimer arrangement is
positionally disordered since, similarly to the monomer-dimer case
illustrated in Fig.~1(a), it is not possible to tile the 
underlying lattice alternately with dimers and trimers.
At a filling of $N_c/N_s=9/4$ we find the ordered state shown 
in Fig.~3(a). Here the trimers form 
a triangular lattice 
where every other substrate minimum in every other row 
captures three colloids. 
In addition, the trimers are orientationally ordered, as
are the dimers surrounding each trimer.

Since the $N_c/N_s=9/4$ dimer-trimer state has two different types
of orientational order, we consider its melting properties to see
whether the orientational ordering can be destroyed in more than one
step.  We find four phases as a function of temperature,
illustrated in Fig.~3.  
We report temperature in units of $T_{m}$,
the temperature at which the colloidal lattice at 
this density melts in the absence of the substrate.  
Phase I, the frozen orientationally ordered
state described above, appears at low $T$, as shown in Fig.~3(a). 
As $T$ increases, the trimers undergo 

\begin{figure}
\center{
\epsfxsize=3.5in
\epsfbox{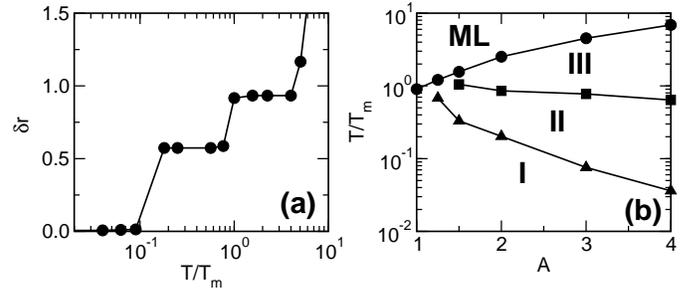}}
\caption{  
(a) The average displacement $\delta r$ versus temperature for the
system in Fig.~3. (b) Phase diagram for
$N_c/N_s=9/4$ at different values of substrate strength $A$,
showing the disordering lines
between the different phases. Triangles: $T_c^{(1)}$, transition
from phase I to phase II;
squares: $T_c^{(2)}$, transition from phase II to phase III; 
circles: dissolution transition from phase III to a modulated liquid (ML).
}
\end{figure}

\noindent
a rotational disordering
transition and begin to rotate at a temperature $T^{(1)}_{c}$; however,
the dimers remain frozen.
In Fig.~3(b) we plot the colloid trajectories (black lines) 
over a fixed time period for this phase, termed phase II, which  
shows that the trimers are undergoing rotation while the dimers are 
stationary and ordered.
As $T$ is further increased, the system enters
phase III at $T^{(2)}_{c}$ when a second orientational disordering transition 
occurs.
Here the dimers lose their orientational order and begin to rotate,
as illustrated in Fig.~3(c),
but the colloids remain trapped within the substrate minima and
there is no diffusion of colloids throughout the system. 
The trajectories of the rotating trimers are widened in Fig.~3(c)
compared to Fig.~3(b) due to the higher temperatures.  
If the temperature is increased sufficiently,
the dimers and trimers dissolve and diffusion occurs throughout the 
whole system in a modulated liquid, as shown in Fig.~3(d).     
Both the dimers and the trimers appear to dissolve at the same temperature, 
so we do not find an interstitial liquid of colloids moving
throughout the sample coexisting with a pinned solid, such as
seen in systems where the substrate has  
flat regions between traps \cite{Leiderer,Vortexmelt}.  

The transitions between the different phases can be 
observed by monitoring the average particle displacements $\delta r$. 
In Fig.~4(a) we plot $\delta r$ vs $T/T_m$ for the system from Fig.~3. 
Here, $\delta r=0$ in phase I.  The transition to the trimer liquid,
dimer solid phase II appears as a step in $\delta r$ at
$T_c^{(1)}$, while a
second step occurs at $T_c^{(2)}$ at the transition to the dimer-trimer liquid
phase III.  In the modulated liquid phase $\delta r$ increases rapidly.
Note that although there are sharp jumps in $\delta r$, these are not
signatures of a first order transition but instead arise due to our
definition of the measure.
We have used 
$\delta r$ to
identify the transitions between the phases 
in a series of simulations for different substrate
strengths $A$ at $N_c/N_s=9/4$. 
We consider only $A > 1$, since 
at this filling for $A<1$ new incommensurate modulated
triangular ordering states appear as the colloid-colloid
interactions become dominant. 
As indicated in Fig.~4(b),
the transition temperatures
$T^{(1)}_{c}$ and $T^{(2)}_{c}$ both decrease as $A$ increases,
with $T_c^{(1)}$ decreasing more rapidly
than $T_c^{(2)}$. 
In contrast, the crossover line
from phase III to the modulated
liquid increases linearly with $A$.                   
In simulation \cite{Reichhardt} and experiments \cite{Bechinger}
on single-species CMCs, the
transition temperature from the oriented CMC state to the orientationally
disordered CMC state decreases with increasing substrate strength, 
as also seen here for the multi-species CMC. 
Theoretical work for the single species case \cite{Agra}
has shown that this is due to the reduction of the
effective multipole moments when the colloids in
each substrate minimum are forced closer together with increasing $A$. 
In phase I of the multi-species case, 
the long range orientational order of the trimers 
arises due to interactions with the other trimers. 
Because the average trimer-trimer spacing is much larger than 
the average dimer-dimer spacing, the trimer orientational
ordering energy is smaller than the dimer orientational energy and
thus the trimers undergo a rotational disordering transition
at a lower temperature than the dimers.  
We note that similar multi-step melting transitions occur for other filling
fractions whenever both $n$- and $m$-mer species
have orientational ordering at low temperatures.

In conclusion, we have shown that in  
colloidal molecular crystals composed of 
monomer-dimer or dimer-trimer mixtures, 
a rich variety of novel ordered and
partially ordered states can occur. 
The ordered monomer-dimer states include  
star and pinwheel states, where the latter are similar to phases
observed for molecular dimers adsorbed on periodic substrates.
Unlike the molecular system, it is possible to observe
the colloidal versions of these phases directly in experiment,
and in addition the strength of the quadrupole interaction between dimers can
be tuned by varying the substrate strength.
For other monomer-dimer fillings, partially ordered or
frustrated states can occur, including a positionally ordered but
orientationally disordered state 
which may offer a convenient realization of an orientational glass.
For ordered states in which the two species both have orientational
ordering, such as dimer-trimer mixtures,
a multi-step melting occurs where one species rotationally disorders while
the other species remains orientationally ordered.
As the temperature increases further,
the second species loses its orientational order 
and finally at high temperatures the $n$-mer states dissociate
into a modulated liquid.
In addition to colloids on periodic substrates, 
similar states may be realizable for vortices
in superconductors with periodic pinning or for
vortices in Bose-Einstein condensates interacting with periodic substrates. 

We thank C.~Bechinger 
for useful discussions. 
This work was supported by the U.S.~Department of Energy under
Contract No.~W-7405-ENG-36.


\end{document}